\pdfoutput=1
\documentclass[prb,twocolumn]{revtex4}   

\usepackage{amsmath}    
\usepackage{graphicx}   

\begin{document}

\title{A simple, low-cost, data-logging pendulum built from a computer mouse}
\author{Vadas Gintautas}
 \altaffiliation[Present address: ]{Center for Nonlinear Studies, Theoretical Division, Los Alamos National Laboratory}  
 \email{vadasg@lanl.gov}   
 \author{Alfred H{\"u}bler}
 \affiliation{Center for Complex Systems Research, Department of Physics, University of Illinois at Urbana-Champaign}
\date{\today}

\begin{abstract}
Lessons and homework problems involving a pendulum are often a big part of introductory physics classes and laboratory courses from high school to undergraduate levels.  Although laboratory equipment for pendulum experiments is commercially available, it is often expensive and may not be affordable for teachers on fixed budgets, particularly in developing countries.  We present a low-cost, easy-to-build rotary sensor pendulum using the existing hardware in a ball-type computer mouse.  We demonstrate how this apparatus may be used to measure both the frequency and coefficient of damping of a simple physical pendulum.  This easily constructed laboratory equipment makes it possible for all students to have hands-on experience with one of the most important simple physical systems.
\end{abstract}

\maketitle

\section{Introduction}
Lessons and homework problems
involving a pendulum are often a big 
part of introductory physics classes and 
laboratory courses.~\cite{Gluck_2004}  Typically 
experiments are limited to using 
photogates to measure the period of the 
pendulum.  Commercial rotary motion 
sensors\footnote{Vernier order number RMS-BTD; PASCO model number PS-2120}
that allow students to collect 
real-time motion data for a pendulum 
exist, but often the cost is too great to 
provide each student in the class with 
such a sensor, especially in developing countries.  
In contrast, a new two-button
button ball-type mouse can be purchased for 
under $5$ US dollars and surplus used units
are often available at little to no cost.
Therefore we present a low-cost, 
easy-to-build rotary sensor pendulum 
using the existing hardware in a 
computer mouse. 

There have been other attempts to use 
common computer peripherals as data 
acquisition interfaces.  T. J. Bensky in 
2001 described the use of a computer 
joystick to track the motion of a 
pendulum.~\cite{Bensky_2001}  We considered using his 
design when building a data-logging 
pendulum, but computer joysticks have 
changed considerably in the past $8$ years.  
Few, if any, models are sold that do not 
self-center; this is a crucial feature to 
Bensky's original design.  Three papers 
by Handler, Ochoa, and Kolp feature the 
use of a computer mouse in tracking 
motion in a Lenz's Law experiment and 
in harmonic motion experiments using 
springs.~\cite{Handler_1996, Ochoa_1997, Ochoa_1998}  In each case a string was 
wrapped around the roller in the mouse 
so that a linear displacement could be 
measured.  

In the last $10-12$ years since these experiments, 
significant changes in mouse hardware 
years have made it easier than ever to use a 
computer mouse to measure angular 
displacement as well.   
At the time that these papers were 
written, the rollers in the mouse featured 
a series of electrical contacts that would 
produce a signal by brushing against a 
wire as the mouse was moved.  This has 
since been nearly universally replaced 
by an opto-mechanical mechanism that 
is much lower in friction.  The rollers 
now have slotted disks, and photogates 
sense the motion of the disk without any 
physical contact required.  This 
improved hardware is ideal for tracking 
the motion of a pendulum because 
measurement friction is low and because modern
ball-type mice feature opto-mechanical mechanisms
which directly measure angular displacement.

\begin{figure}
    \includegraphics[width=0.9\linewidth]{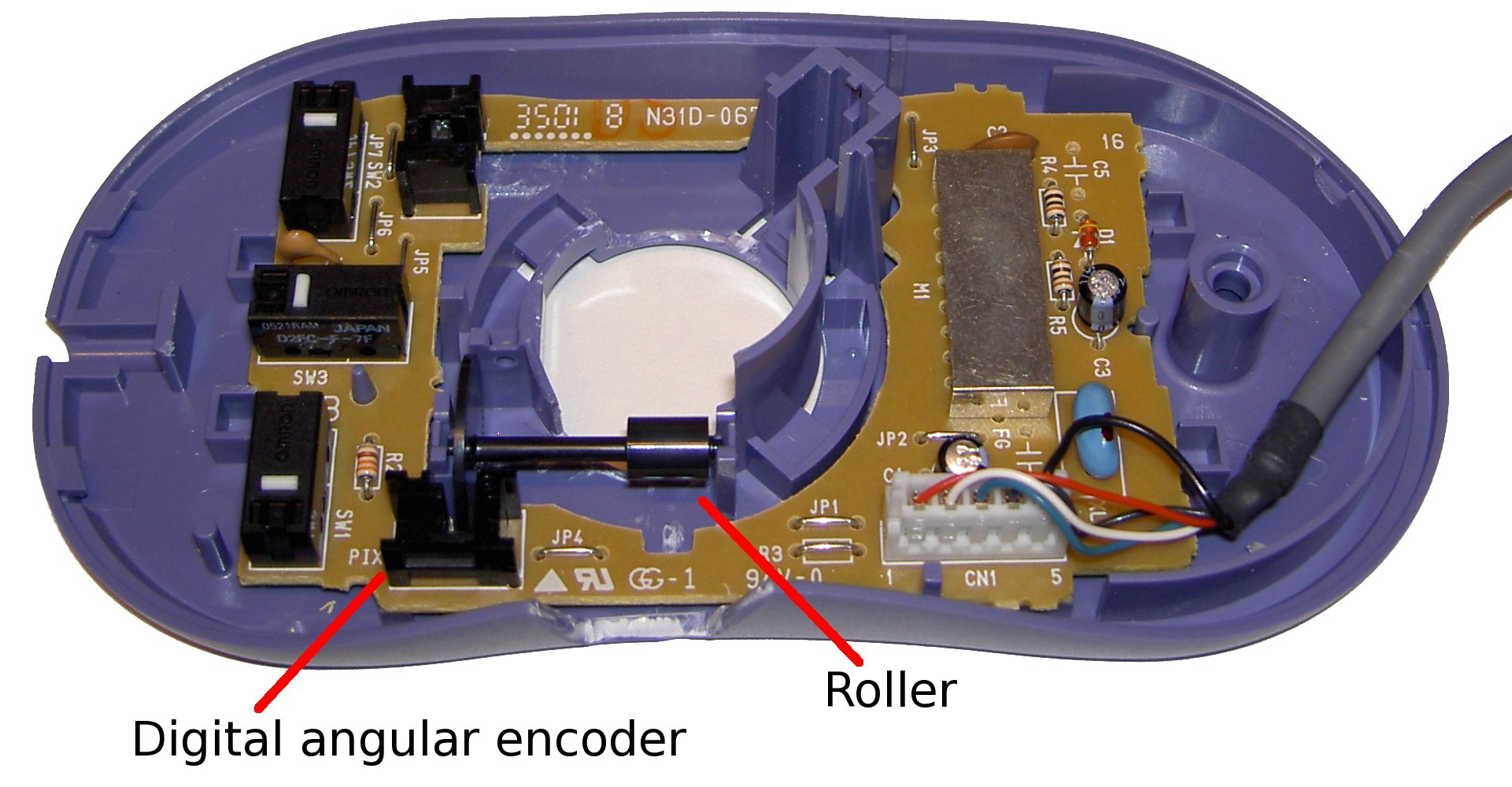}
    \caption{Interior of a computer mouse.  The hole in the center is for the ball, which has been removed.  The key components are the digital angular encoder and the associated roller, as indicated.  The pendulum will be attached directly to this roller.  The second roller (to detect translation orthogonal to the first roller) has been removed.  Color online.}
    \label{fig:mouse}
\end{figure}

There exist computer mice that use a purely
optical mechanism to detect translation, namely,
a low resolution camera on the underside of the
mouse.  The camera repeatedly takes pictures of 
the surface under the mouse and interprets differences
between successive frames as motion.  While these mice
work the same way to the user, this type of mouse has no 
mechanical components and is not well suited for this
application.  It may be possible to adapt an optical
mouse for other mechanical experiments, but in this work
we specifically take advantage of the hardware present in
ball-type mice.

\section{Building the pendulum}
The following parts and tools are needed: a ball-type computer
mouse, a wooden or plastic dowel, a small screwdriver (used to
open the mouse casing), a pair of small clippers (used to cut
back the mouse casing), and a small drill bit (approximately the diameter
of the dowel).
We remove the cover of the mouse 
and locate the best digital angular encoder (with roller) to 
use for a pendulum (see Fig.~\ref{fig:mouse}).  We cut 
away enough of the cover to allow 
access to the encoder, then replace the 
cover to provide support to the assembly.  
In this case we allow for moderate 
amplitude ($180^\circ$) motion of the 
pendulum by cutting back the plastic near
the roller.  With additional hardware it 
is possible to mount the brackets of the 
roller from either side
to allow for full $360^\circ$ motion
of the pendulum.  If 
the pendulum is not mounted directly to the 
roller but the pendulum is suspended independently
and the roller is connected directly to it,
measurement friction can be eliminated. 

\begin{figure}
    \includegraphics[width=0.9\linewidth]{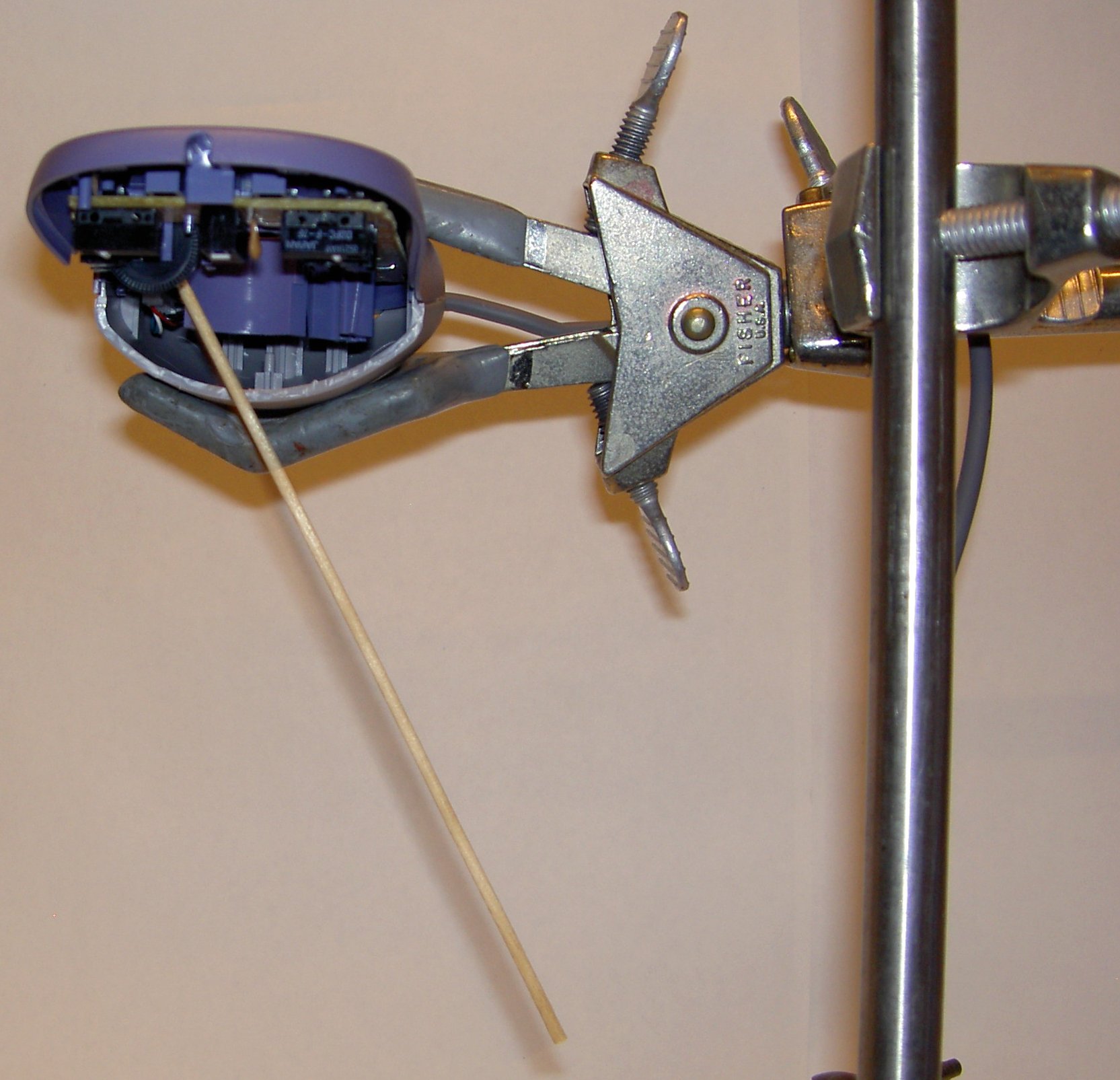}
    \caption{The assembled pendulum-mouse in action.  Color online.}
    \label{fig:pend}
\end{figure}

A small drill bit turned by hand will 
make a hole in the roller.  A rod with the 
same diameter as the drill bit will fit into 
this hole and will function as the 
pendulum.  We used a thin wooden 
dowel,\footnote{This pendulum has a length of $14.7$cm, a diameter of $2.2$mm and a mass of $0.38$g.} but a small plastic rod 
would work equally well.  Since no glue 
is used, rods of varying lengths can be 
easily substituted during experiments.  Fig.~\ref{fig:pend}
 shows the completed experimental 
apparatus. 
 
\section{Calibration and Use}
We plug the mouse into the Universal Serial Bus (USB)
port of a computer.\footnote{In general, most operating systems will accept input from two mice plugged in 
simultaneously (via the USB port) so the 
pendulum need not interfere with normal 
operation of the computer.  
This was tested for Windows XP, Linux, and OS X.
Mice that use other ports such as serial or PS/2 may also
work in this way but nearly all computers built
within the last 10 years feature 
multiple USB ports so this is most convenient.
Be sure to turn OFF the hardware acceleration for the mouse to ensure that the displacement does not depend on the angular velocity of the roller.}
The resolution of the apparatus is limited by 
the number and spacing of the slots in 
the disk of the angular encoder.  
Therefore, it is possible to calibrate the 
motion of the cursor to angular 
displacement units.  One calibration 
method is to determine the motion in 
pixels of one or more full turns of the 
roller.  For the apparatus we built, a rotation of $360  
\pm 1$ degrees gave a change in the position 
of the cursor of $194 \pm 1$ pixels.  This 
results in a conversion factor of $0.0324$ 
radians per pixel.  A custom computer program is used to 
the motion of the cursor in
pixels.\footnote{It is somewhat more difficult
but quite possible to write a computer program
that uses the data coming
from the mouse itself.  The resolution of this
data is limited only by the resolution of the optical angular 
encoder within the mouse.  See Endnote $13$ for a link to the 
source code of the computer program used to track the cursor.}
We find that the apparatus has a resolution 
of $0.0324$ radians, which corresponds to 
a fractional uncertainty of $0.52\%$ out of 
a full turn. 

The standard equation of motion of a 
pendulum with damping is  
\begin{equation}
    Ia = -\frac{mgl}{2}\sin{x}-\beta v,
\end{equation}
where $I=\frac{1}{3}ml^{2}$ is the moment of inertia.
Here $m$ and $l$  refer to the mass and length of the rod, 
respectively,   while $g$ is acceleration due to gravity 
and $\beta$ controls the strength of the 
damping term.  Also, $x$, $v$, and $a$ are the angle 
measured down from the vertical, the 
angular velocity, and the angular acceleration, 
respectively.  In the small angle 
approximation, the equation of motion 
reduces to 
\begin{equation}
    a = -\frac{3g}{2l}x-\frac{3\beta}{ml^{2}} v.
\end{equation}
We can write the solution as follows: 
\begin{equation}
    x(t) = Ae^{-\delta t}\cos{(\omega t _ \phi)},
\end{equation}
where $\delta = \frac{3\beta}{2ml^{2}}$ is the 
decay constant and $\omega = \sqrt{\frac{3g}{2l}-\delta^{2}}$ is the 
frequency of free oscillations.~\cite{Graf_2005}

\begin{figure}
    \includegraphics[width=0.9\linewidth]{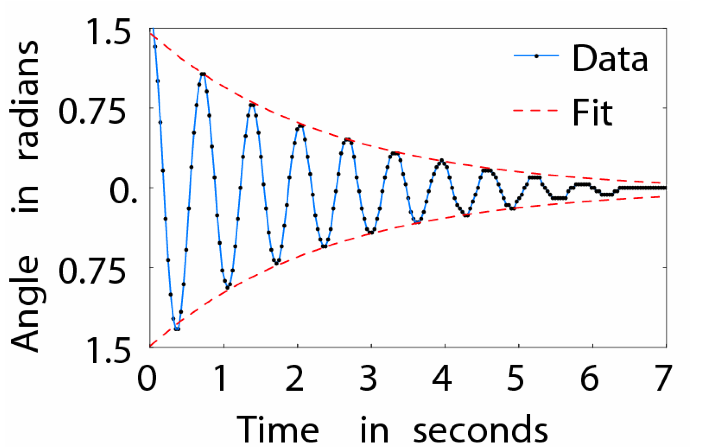}
    \caption{Position-vs-time data for the pendulum.  The dots show the actual data from the 
    pendulum, with a thin solid line added to guide the eye.  The dashed line shows a fit of the data to an exponential decay envelope.  Color online.}
    \label{fig:data}
\end{figure}

$A$ and $\phi$ are determined using the initial 
conditions.  Fig.~\ref{fig:data} shows a plot of 
position versus time data for the 
pendulum when released from rest 
 ).  The dashed line shows a fit of the 
turning points to an exponential decay 
envelope.  For the pendulum, we obtain 
$\delta=0.415$  with $R^{2}=0.997$ for the fit.  
This simple investigation is quite easy to 
do using the data from this apparatus, but 
next to impossible using only photogates.
This is a simple example of a high school or 
undergraduate level laboratory experiment that 
uses the apparatus but it is well suited to more
advanced teaching applications such as
exploring the driven physical pendulum 
or coupling between pendula.
   
\section{Conclusion} 
The mouse-pendulum is a low cost solution 
to the need for an experimental 
pendulum that provides real-time 
angular displacement data.  The small-amplitude
period of the pendulum (approximately $0.63$s 
for the apparatus we built) depends only on the 
length of the rod and can be easily 
adjusted for various experiments.  This 
apparatus is ideal for undergraduate and 
even high school lab classes.  
Furthermore, the low cost makes it possible for
teachers with very limited budgets and those in developing countries
to provide each student in the class with a useful piece of lab equipment.
The ease of construction allows building the pendulum to be 
a good classroom exercise, during which practical experimental
considerations (such as how to minimize friction) may be discussed.
This design is robust for serious experimental work as well.\footnote{A variation of
this design was successfully used to explore synchronization between real and virtual pendula experimentally.~\cite{Gintautas_2007}}
Source code to free data logging software to record the motion of the pendulum can be obtained online.\footnote{\texttt{http://vadasg.googlepages.com/PendDataLog-src.zip}}  LA-UR 09-00439.


\end{document}